\documentclass[showpacs,onecolumn,superscriptaddress]{revtex4}

\usepackage{doi}%<----------
\usepackage{hyperref}
\hypersetup{
  colorlinks=true,        
  linkcolor=blue,         
  citecolor=cyan,         
}

\usepackage{graphicx}
\usepackage{xcolor, soul}
\sethlcolor{green}
\usepackage{dcolumn}
\usepackage{bm}
\usepackage{color}
\usepackage{enumitem}
\usepackage{mathrsfs}
\usepackage{amsmath}
\usepackage{amssymb}
\usepackage{cleveref}

\begin{document}

\title{Overcharging process around a magnetized black hole:\\ Can the backreaction effect of magnetic field restore cosmic censorship conjecture? }

\author{Sanjar Shaymatov}\email{sanjar@astrin.uz}
\affiliation{Institute of Fundamental and Applied Research, National Research University TIIAME, Kori Niyoziy 39, Tashkent 100000, Uzbekistan}
\affiliation{Akfa University,  Milliy Bog Street 264, Tashkent 111221, Uzbekistan}
\affiliation{National University of Uzbekistan, Tashkent 100174, Uzbekistan} 
\affiliation{Tashkent State Technical University, Tashkent 100095, Uzbekistan}
\affiliation{Samarkand State University, University Avenue 15, Samarkand 140104, Uzbekistan}

\author{Bobomurat Ahmedov}
\email{ahmedov@astrin.uz}
\affiliation{Ulugh Beg Astronomical Institute, Astronomicheskaya 33, Tashkent 100052, Uzbekistan}
\affiliation{Institute of Fundamental and Applied Research, National Research University TIIAME, Kori Niyoziy 39, Tashkent 100000, Uzbekistan}
\affiliation{National University of Uzbekistan, Tashkent 100174, Uzbekistan}

%date{\today}

\begin{abstract}
It is well known that the electrically charged Reissner-Nordstr\"{o}m black hole could be overcharged. Here, we investigate 
the process of overcharging of a magnetized Reissner-Nordstr\"{o}m  black hole that includes effect of the magnetic field generated by own magnetic charge of source on the background geometry. It is found that magnetic field prevents a transition to occur from black hole to naked singularity, thus overcharging cannot be attained which happens due to the fact that  
the magnetic field reaches its threshold value. It turns out
that beyond threshold value the magnetic field can exert  
large Lorentz force on particles and dominate over the gravitational force, allowing charged particles not to fall into the black hole. One may conclude, there occurs no evidence for violation of cosmic censorship conjecture for a magnetized Reissner-Nordstr\"{o}m  black hole beyond threshold value of the magnetic field. 
\end{abstract}
%\pacs{04.70.Bw, 04.20.Dw}
\maketitle

%\section{Introduction}

\section{Introduction}
\label{Introduction}

\textcolor{black}{In general relativity (GR) the singularity theorems was first proposed by Penrose in 1965 in pioneering paper~\cite{Penrose65a}, implying that the singularities likely occur in case  matter obeys certain energy conditions. Later, this theorem was extended to investigate the conditions as to the emergence of singularities in GR, i.e. called Penrose-Hawking theorem~\citep{Hawking-Penrose70}. The occurrence of singularities is a breakdown of Einstein's theory because of their geodesic incompleteness. Irrespective of the fact that the Penrose-Hawking theorem has given the evidence in favour of the existence of singularities in GR, it did not shed light on the properties of the singularities. Thus, the cosmic censorship hypothesis in the weak form was proposed by Penrose \citep{Penrose69} in 1969 in order for the Einstein gravity to keep valid, thus hiding the singularity from being seen for outside observers. The validity of the weak cosmic censorship conjecture (WCCC) would not make the singularities possible to observe the final state of a sufficiently compact massive object as a result of gravitational collapse. Consequently, black holes are very intriguing objects as a generic solution  of the field equations of GR as well as with their remarkable geometric nature. It is worth noting here that recent  observational studies of gravitational wave astronomy \cite{Abbott16a,Abbott16b} and supermassive black hole that exists at the center of the elliptical  galaxy Messier 87  (M87) through imaging by the Event Horizon Telescope (EHT) and BlackHoleCam ~\cite{Akiyama19L1,Akiyama19L6} have provided solid and trustful information in strong gravity regime that verifies the existence of black holes in nature. These gravitational wave and sub-millimeter radiowave observations (together with infrared one around Sgr A* at the center of our galaxy)  provide strong and very potent tests to understand deeply the unknown aspects of black holes, yet there still exist open questions associated with the occurrence of physical  singularity which marks the limit of classical Einstein's theory applicability. In this regard, WCCC can be considered as the main tool in testing GR. 
} 

\textcolor{black}{Wald first proposed the formulation of the validity of the  WCCC to destroy the black hole horizon by test particles~\cite{Wald74} and it was shown that the WCCC can not be violated for extremal black hole. Later, such a process was extended by somewhat different prospective, according to which it is not possible for test particle with appropriate parameters to reach the black hole horizon as there is no parameter space~\cite{Dadhich97}. However, Hubeny addressed this issue somewhat differently~\cite{Hubeny99}, accordingly  showing the possibility of turning a nearly extremal black hole into a naked singularity via charged test particles with  appropriate parameters. The above process was extended to the rotating black holes, i.e. it was shown that Kerr/Kerr-Newman black hole could be overspun/overcharged in case when a falling in particle adds enough angular momentum/charge to the black hole's angular momentum/charge~\cite{Jacobson09,Saa11}. Numerous papers have since been devoted to the study of testing the WCCC in this context in various gravity models \citep[see, e.g.][]{Rocha14,Shaymatov15,Song18,Duztas18,Gwak18a,Jana18,Shaymatov19a,Duztas-Jamil18b,Duztas-Jamil20,Jiang20,Yang20a,Shaymatov22JCAP}. Later on, it was shown that the WCCC could be held if and only if self-force effects are included~\cite{Barausse10,Li13,Zimmerman13,Rocha11,Isoyama11,
Colleoni15a,Colleoni15b}. The above analysis was also extended to the variety of situations, i.e. for BTZ black holes \cite{Duztas16}, black hole with charged scalar field \cite{Gwak20}, black hole dynamics~\cite{Mishra19} and higher dimensions \cite{Bouhmadi-Lopez10,Revelar-Vega17}. Recently, the above thought experiment has been developed by Sorce and Wald \cite{Sorce-Wald17,Wald18}, thus referring to the new gedanken experiment including the nonlinear particle accretion process always favoring the validity of the WCCC. Following Sorce and Wald there has been a extensive body of research work~\cite{An18,Ge18,Ning19,Yan-Li19,Jiang20plb,
Shaymatov19c,Shaymatov20a,Shaymatov21a,Shaymatov21d} addressing the question of overcharging/spinning of black hole under the nonlinear order perturbations for $D\geq4$ dimensions. }

\textcolor{black}{From astrophysical point of view, it is believed that a test magnetic field may exist in the environment surrounding the black hole. With this in view, the magnetic field could play a decisive role in altering the geodesics of charged test particles. There was investigation that explores the effect of the external magnetic field on the particle geodesics to test whether it could violate the WCCC~\cite{Shaymatov15}. It was shown that the magnetic field can serve as a cosmic censorship conjecture beyond its certain critical value, thus affecting on the particle geodesics drastically and preventing particles from reaching the black hole horizon as that of large Lorentz force. It happens because the magnetic field backreaction on the background geometry must be slightly stronger as compared to the small backreaction induced by the test particle. This was true for the external magnetic field, however, what happens provided that the charged black hole is magnetized one, i.e. does it still act as a cosmic censorship conjecture? To settle this question we use the magnetized black hole solution that includes the magnetic field in the background spacetime~\cite{Gibbons13}. Note that the inclusion of the magnetic field backreaction on the background spacetime is impossible due to the fact that there is no exact solution. However, the solution describing the magnetized Reissner-Nordstr\"{o}m solution has recently been derived in Ref.~\cite{Gibbons13}. In this paper, following~\cite{Gibbons13} we investigate the magnetized black hole to understand more deeply the backreaction effect of the magnetic field on the validity of the WCCC, thus leading to reach the definite conclusion for the magnetic field. Interestingly we show that the magnetic field could still serve as a cosmic censor, i.e. the black hole can never be overcharged beyond the critical value of the magnetic field.  }

\textcolor{black}{In realistic astrophysical scenario, it is particularly important to understand completely the impact of the existing fields on the particle geodesics in the environment surrounding the black holes. Of them the magnetic field is increasingly important to explain rich astrophysical phenomena around black holes. For example, the magnetic field can influence the motion of charged particles drastically and can alter the particle geodesics. There have been numerous  works ~\cite{Frolov10,Aliev02,Shaymatov14,Shaymatov18a,Dadhich18,Shaymatov19b,
Shaymatov20egb,Shaymatov21c,Haroon19,Konoplya19plb,Hendi20,Jusufi19,Shaymatov21-b,Narzilloev20a,Rayimbaev-Shaymatov21a,Shaymatov21pdu,Shaymatov22b} addressing the impact of the magnetic field on the  particle motion in a variety of situations. The magnitude of  the magnetic field $B$ is gauged to be of order $\approx 10^8~G$ for stellar mass black holes and $\approx 10^4~G$ for supermassive black holes, respectively (see for example \cite{Piotrovich10}). Also there is another way that can be considered to estimate its magnitude at the black hole horizon radius~\cite{Eatough13,Shannon13}, and it was later shown that the magnetic field is estimated to be between $200~G$ and $8.3 \times 10^{4}~G$ at 1 Schwarzschild radius~\cite{Baczko16}. Recent advances in infrared, optical, x-ray, and radio observational studies of binary black holes system V404 Cygni has provided that the magnitude of the magnetic field would be of $33.1 \pm 0.9 G$ (see  for example \cite{Dallilar2018}).   }

In Sec.~\ref{Sec:magnetized}, we describe briefly the magnetized Reissner-Nordstr\"{o}m black hole and charged particle motion which is followed by the main study of dynamics of overcharging of the magnetized Reissner-Nordstr\"{o}m black hole~in Sec.~\ref{Sec:overcharging}. We end up with a conclusion in  Sec.~\ref{Sec:Conclusion}. Throughout the manuscript we use a geometric system of units in which $G=c=1$.

\section{Magnetized Reissner-Nordstr\"{o}m black hole and particle motion }\label{Sec:magnetized}

The spacetime metric describing a magnetized charged Reissner-
Nordstr\"{o}m black hole in Schwarzschild coordinates ($t,r,\theta, \phi$) is given by~\cite{Gibbons13}
\begin{eqnarray}\label{Eq:metric} 
d s^2 &=& H\, [-f dt^2 + f^{-1}\, dr^2  + r^2 d\theta^2] +
      H^{-1}\, r^2\sin^2\theta\,(d\phi -\omega dt)^2\, ,
\end{eqnarray}
where 
\begin{eqnarray}
f&=& 1- \frac{2M}{r} + \frac{Q^2}{r^2}\, ,\\ 
\omega &=& -\frac{2Q B}{r} + \frac{1}{2} Q B^3\, r (1+f
\cos^2\theta)\, ,\\
H &=& 1 +\frac{1}{2}B^2 (r^2\sin^2\theta + 3 Q^2\cos^2\theta)
+  \frac{1}{16} B^4 (r^2 \sin^2\theta + Q^2\cos^2\theta)^2\, ,
\end{eqnarray}
with $M$ and $Q$ which, respectively, refer to the black hole's total mass and charge. Note that the parameter $B$ is related to the magnetic field. 
With $Q$ and $B$ given for the magnetized charged black hole vector potential for the electromagnetic field is written as follows 
\begin{eqnarray}
A = A_t dt + A_\phi (d\phi-\omega dt)\, ,
\end{eqnarray}
with the following electromagnetic vector potential components 
 \begin{eqnarray}
A_t &=& -\frac{Q}{r} +\frac{3}{4} Q B^2 r\, (1+ f\cos^2\theta)\,,\nonumber\\
A_\phi &=& \frac{2}{B} - H^{-1}\Big[\frac{2}{B} +
             \frac{1}{2} B(r^2\sin^2\theta + 3 Q^2
             \cos^2\theta)\Big]\, .
\end{eqnarray}
%%%%%

\textcolor{black}{Further we focus on a charged particle's motion around magnetized Reissner-Nordstr\"{o}m black hole, for which the Hamiltonian of the system is defined by  \cite{Misner73}
\begin{eqnarray}
H=\frac{1}{2}g^{\mu\nu} \left(\pi_{\mu}-qA_{\mu}\right)\left(\pi_{\nu}-qA_{\nu}\right)\, ,
\end{eqnarray}
with $\pi_{\mu}$ and $A_{\mu}$ referred to as a charged particle's canonical momentum and the four-vector potential of the electromagnetic field. In the Hamiltonian, the charged particle's four-momentum is given as
\begin{eqnarray}
p^{\mu}=g^{\mu\nu}\left(\pi_{\nu}-qA_{\nu}\right)\, . 
\end{eqnarray}
Hence, one can then write the equations of motion for the charged particle as 
\begin{eqnarray} 
\label{Eq:H1}
  \frac{dx^\alpha}{d\lambda} &=& \frac{\partial H}{\partial \pi_\alpha}\, ,  \\
  \frac{d\pi_\alpha}{d\lambda} &=& - \frac{\partial H}{\partial x^\alpha}\, , \label{Eq:H2}
\end{eqnarray}
with $\lambda=\tau/m$ being the affine parameter associated with the proper time $\tau$ for
timelike geodesics. }

\textcolor{black}{From Eqs.~(\ref{Eq:H1}) and (\ref{Eq:H2}) it is then  straightforward to obtain the equations of motion for the charged particle. 
According to the symmetry property of the magnetized black hole spacetime
admitting two Killing vectors,
$\xi^{\alpha}_{(t)}=(\partial/\partial t)^{\alpha}$ and
$\xi^{\alpha}_{(\phi)}=(\partial/\partial \phi)^{\alpha}$ being
responsible for stationarity and axisymmetry, the energy and
angular momentum of the charged particle are determined by
\begin{eqnarray}
\label{En1} -\delta E&=& -g_{\mu\nu}(\xi_{t})^{\mu}\pi^{\nu}=
g_{tt}\pi^{t} + g_{t\phi}\pi^{\phi}+qA_{t},\\
 \label{Ln1}
\delta J&=& -g_{\mu\nu}(\xi_{\phi})^{\mu}\pi^{\nu}=g_{\phi t}\pi^{t} +
g_{\phi\phi}\pi^{\phi}+qA_{\phi},
\end{eqnarray}
where $\pi^{\nu}$ is the four velocity defined by
$\pi^{\nu}=\frac{dx^{\nu}}{d\tau}$ with the proper time $\tau$ for
timelike geodesics. Here we note that the system represents four independent constants of motion of which we have specified three, i.e., $\delta E$, $\delta J$ and $m^2$. The other constant is related to the latitudinal particle motion. However, in the following, we will restrict ourselves to the equatorial motion, setting $\theta=\pi/2$ and therefore ignoring the fourth constant of motion.}

%%%%%%%%%%%%%%%%%%%%%%%%%%%%%%%%%%%%%%%%%%%%%%%%%%%%%%%%%5
\textcolor{black}{Taking into account Eqs. (\ref{Eq:H1}-\ref{Ln1}) with the normalization condition $g_{\mu\nu}p^{\mu}p^{\nu}=-m^2$, we write the radial part of the equation of motion for massive particles in the following form 
\begin{eqnarray}
\frac{1}{2}\dot{r}^{2} +V_{\rm
eff}\left(r\right) &=& 0,
\end{eqnarray}
where the effective potential
for radial motion of charged particle, $ V_{\rm
eff}\left(r\right)$, is given by \cite{Shaymatov21c}}
%
%\begin{widetext}
\begin{eqnarray}\label{Eff1}
V_{\rm
eff}\left(r\right)&=&\left(
fH^{-1}-\frac{Hf^{-1}}{m^2r^2\left(\omega^2r^2-fH^2\right)}\left[q^2\left(\frac{\omega^2r^2}{H}-f
H\right)\left(\frac{2}{B}-H^{-1}\left[\frac{2}{B}+\frac{1}{2}Br^2\right]\right)^2\right.\right.\nonumber\\&+&\frac{2~q
\omega
r^2}{H}\left(\frac{2}{B}-H^{-1}\left[\frac{2}{B}+\frac{1}{2}Br^2\right]\right)
\left(\delta E-\frac{q Q}{r}+\frac{3}{4}q Q
B^2r\right)\nonumber\\&+&\left.\left.\frac{r^2}{H}\left(\delta E-\frac{q Q}{r}+\frac{3}{4}q Q
B^2r\right)^2\right]\right) \, .
\end{eqnarray}
%\end{widetext}
%\textcolor{red}{}

In the limiting case when magnetic field vanishes $B\rightarrow 0$ one can recover Hubeny's result for effective potential~\cite{Hubeny99}  
%
%\begin{widetext}
\begin{eqnarray} \label{Hub_eff} V_{eff}(r)=
\frac{M}{r}\left(\frac{Q}{M}\frac{{\delta E}
q}{m^2}-1\right)-\frac{1}{2}\left(\frac{{\delta E^2}}{m^2}-1\right)
-\frac{Q^2}{2r^2}\left(\frac{q^2}{m^2}-1\right)\ .
\end{eqnarray}
%\end{widetext}}

\section{Dynamics of overcharging of the magnetized 
Reissner-Nordstr\"{o}m black hole } \label{Sec:overcharging}

Here, we test the validity of the WCCC in the case of charged particles interacting with the magnetized  Reissner-Nordstr\"{o}m black hole by applying the gedanken experiments. This is what we wish to address in this section.

\subsection{Overcharging of Reissner-Nordstr\"{o}m black hole }

Now we first consider the Reissner-Nordstr\"{o}m black hole immersed in the external magnetic field. Here the main purpose is to ensure that whether the charged particle with appropriate parameters can reach the black hole horizon in the presence of external magnetic field, thereby violating the WCCC. Note that the horizon radius for Reissner-Nordstr\"{o}m black hole is given by $r_{\pm}=M+\sqrt{{M}^2-Q^{2}}$. In this regard, the horizon stability must be hold provided that $M>Q$ is satisfied, whereas for $M<Q$ the horizon no longer exists. To reach the latter the radially falling charged particle should carry mass $m$ and charge $q$ to the Reissner-Nordstr\"{o}m black hole. \textcolor{black}{Here, it is worth noting that we suppose $\delta E\ll M$ and $q \ll Q$ for test particle approximation to hold good so that the radially falling charged particle transfers the mass and charge to black hole's mass and charge, respectively.} With the mass and charge absorbed, the final state of the black hole parameters is given by $M+\delta E$ and $Q+q$, respectively. 

Thereafter, the black hole can be overcharged when the condition below is satisfied only:
\begin{eqnarray}\label{Eq:11}
(\bar{M}+\delta E)^2<(\bar{Q}+q)^2\, ,
\end{eqnarray}
for the lower and an upper bounds on the energy of the charged
particle
\begin{eqnarray}\label{lower} \frac{q Q}{r_{+}^2}=\delta E_{min}< \delta E\, ,\\
\label{upper} \delta E < \delta E_{max}= \bar{Q}+q-\bar{M}\,.
\end{eqnarray}
Note that the black hole starts out very close to extremal one.
It is certain that the above Eqs.~(\ref{lower}) and (\ref{upper}) cannot be satisfied simultaneously for the extremal case, $Q=M=r_{+}$. 

The effective potential for radial motion of the radially falling charged particle can be given in terms of the energy, charge and mass of the charged particle, as given by Eq.~(\ref{Hub_eff}). 
This effective potential has turning points at which $V_{\rm
eff}(r)=0$. However, it also has maximum at
\begin{eqnarray} \label{rmax} r_{\rm max}=
\left[Q\left(\frac{q^2}{m^2}-1\right)\left(\frac{\delta E
q}{m^2}-\frac{M}{Q}\right)^{-1}\right]^{1/2}\, . \end{eqnarray}
For this maximum one must choose reasonable parameters of charged test particle so that $V_{\rm eff}(r)<0$, thus allowing the particle to cross through the horizon swimmingly. \textcolor{black}{For this purpose
we select the appropriate parameters of the charged particle to satisfy the condition (see Eq.~(\ref{Eq:11})) and explore $V_{\rm eff}(r)$ numerically~\cite{Hubeny99}}
\begin{eqnarray} \label{abs}
Q&=&1-2\epsilon^2 \nonumber\\
q&=&\alpha^{\prime} \epsilon \mbox{~~with~~} \alpha^{\prime} > 1 \nonumber\\
\delta E &=&\alpha^{\prime} \epsilon -2\beta^{\prime} \epsilon^2
\mbox{~~with~~ } 1<\beta^{\prime}
<\alpha^{\prime} \nonumber\\
m&=&\gamma^{\prime} \epsilon \mbox{~~with~~ } \gamma^{\prime}
<\sqrt{\alpha^{\prime 2}-\beta^{\prime 2}} \, . \end{eqnarray}
with suitable choice of parameters $\alpha^{\prime}$,
$\beta^{\prime}$, and $\gamma^{\prime}$. \textcolor{black}{In the above equation, $\epsilon$ is regarded as a small parameter, so that it represents a near-extremality. We note that parameters $\alpha^{\prime}$,
$\beta^{\prime}$, and $\gamma^{\prime}$ have no particular meanings, but are used to satisfy the above-mentioned condition; i.e. $(\bar{M}+\delta E)^2<(\bar{Q}+q)^2$. To ensure this we explore it numerically. For this thought experiment, we can choose $Q=0.99999$ for the given value $\epsilon=0.0022$ by setting $M=1$. One can however choose even smaller values of $\epsilon$. So for the given
background spacetime with parameters $M=1$ and $Q=0.99999$ as
stated in \cite{Hubeny99}, we can then select the appropriate parameters of the charged particle by setting $\alpha'=1.33902$, $\beta'=1.0301$, and $\gamma'=0.80505$. In doing so, we suppose that the particle of mass $m=1.8\times10^{-3}$ has charge $q=3\times10^{-3}$, and falls past the horizon with energy $\delta E=2.9897\times 10^{-3}$, thus satisfying the condition $(\bar{M}+\delta E)-(\bar{Q}+q)<0$.} 

Let us then consider the charged particle thrown inward the black hole radially with $\delta J=0$. For
charged particles to enter the black hole without any restriction
$V_{\rm eff}(r)$ must be negative or $\dot{r}^2$ must
be positive everywhere outside the horizon, thereby converting a near extremality to over extremality. For this, we analyse the radial motion of charged test particle in terms
of the effective potential, $\dot{r}^2=-2 V_{\rm eff}(r)$.
\begin{figure}
\begin{center}
\includegraphics[width=0.5\textwidth]{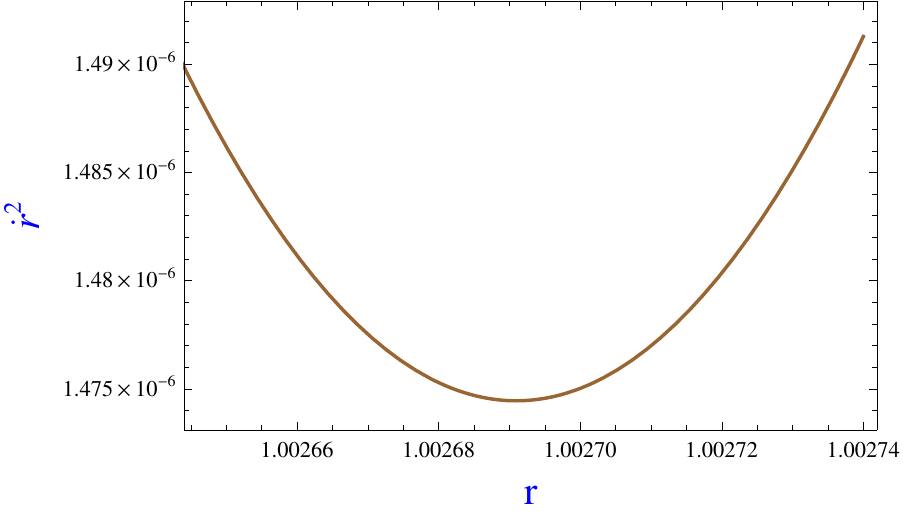}
\caption{\label{figure2} \textcolor{black}{Radial dependence of the motion of
charged particle falling into the near-extremal 
Reissner-Nordstr\"{o}m black hole for given values of charge
$q=3\times10^{-3}$, energy $\delta E=2.9897\times10^{-3}$, and mass $m=1.8\times10^{-3}$ that satisfy the test particle approximation.} }
\end{center}
\end{figure}
In Fig.~\ref{figure2}, we show the radial motion of charged test particle. As seen in Fig.~\ref{figure2}, the charged particle with appropriate parameters could fall across the horizon, thus violating the WCCC -- so-called "over extremality" can be reached.
\begin{table}[ht]
\caption{\label{table1} The values of $r_{\rm max}$ outside the
horizon and effective potential are tabulated for different
values of energy and mass of test particles. The charge parameter is considered here 
as $q=3\times 10^{-3}$.
\\} \centering
\begin{tabular}{c c c c}
\hline\hline
$m$ &~ $\delta E$  &~ $r_{max}$ &~ $V_{eff}$  \\
[0.5ex]
$0.9\times10^{-3}$  &~ $2.9874\times10^{-3}$   &~1.00231   &~$-5.23\times10^{-7}$  \\
[0.5ex]
$1.4\times10^{-3}$   &~ $2.9884\times10^{-3}$ &~ $1.00248$ &~ $-7.87\times10^{-7}$ \\
[0.5ex]
$1.8\times10^{-3}$   &~$2.9897\times10^{-3}$ &~$1.00269$ &~$-1.47\times10^{-6}$   \\
[0.5ex]
$1.85\times10^{-3}$   &~ $2.9899\times10^{-3}$ &~ $1.00273$ &~ $-1.60\times10^{-6}$ \\
[1ex] \hline\hline
\end{tabular}
\end{table}

In Table~\ref{table1} we show the maximum values of the effective potential in a variety of test particle parameters. For these parameters, we explore the effective potential numerically. As can be seen from Table~\ref{table1}, $V_{eff}<0$ is always satisfied for chosen values, thus allowing test particle to fall across the horizon and violating the WCCC. This result also reflects the behavior of Fig.~\ref{figure2} for a particular case. 

\subsection{Effect of external magnetic field on dynamics of overcharging of Reissner-Nordstr\"{o}m black hole }

Now we consider the Reissner-Nordstr\"{o}m black hole immersed in an external magnetic field to understand more deeply the impact of the external magnetic field on the dynamics of overcharging process. The magnetic field can alter the geodesics of charged particles, and it can also influence the process of overcharging of the black hole.
Following to Wald \cite{Wald74} one can study the magnetic field in vicinity of the black hole and assume that the magnetic field is asymptotically uniform with strength $B$ at infinity. The non-vanishing components of potential
$A_{\alpha}$ of the electromagnetic field around the Reissner-Nordstr\"{o}m black hole in the presence of external magnetic field reads as 
\begin{eqnarray}
A_{t}&=&-\frac{Q}{r}\, ,\nonumber\\
A_{r}&=& A_{\theta}=0,\nonumber\\
A_{\varphi}&=&\frac{B}{2}r^2 \sin^2\theta\, .
\end{eqnarray}

We further analyze  effective potential for the radial motion. Using Eqs.~(\ref{En1}) and (\ref{Ln1}) and setting $\theta=\pi/2$ the effective potential describing the radial motion for the charged particle around the Reissner-Nordstr\"{o}m black hole immersed in an external magnetic field can be written in the following form 
\begin{eqnarray} \label{RN_eff}V_{eff}(r)&=&
\frac{M}{r}\left(\frac{Q}{M}\frac{{\delta E}
q}{m^2}-1\right)-\frac{1}{2}\left(\frac{{\delta E^2}}{m^2}-1\right)\nonumber\\
&-&\frac{Q^2}{2r^2}\left(\frac{q^2}{m^2}-1\right)+\left(1-\frac{2M}{r}+\frac{Q^2}{r^2}\right)\frac{\beta^2r^2}{8M^2}\,
,\nonumber\\
\end{eqnarray}
with the magnetic parameter $\beta=qBM/m$. In the limit when $\beta\rightarrow 0$, one can easily recover Eq.~(\ref{Hub_eff}), i.e. the result presented in Ref.~\ref{Hub_eff}. We now analyze the impact of the external magnetic field on the radial profile of the effective potential.

By imposing the following conditions
\begin{eqnarray}
 && \frac{\partial V_{eff}(r)}{\partial r}=0 \mbox{~~and~~ }
V_{eff}(r_{max})=0\, ,
\end{eqnarray}
one can easily find the critical value for the magnetic field at which it prevents the charged particles from falling into the black hole. Taking this in consideration, with the values chosen for the charged test particle,
$q=3\times10^{-3}$ and $m=1.8\times10^{-3}$, one can get the critical value for the magnetic parameter $\beta$ given by
\begin{equation} 
\beta_{cr}\sim 0.7198\, .
\end{equation}

\begin{figure}
\centering
\includegraphics[width=0.5\textwidth]{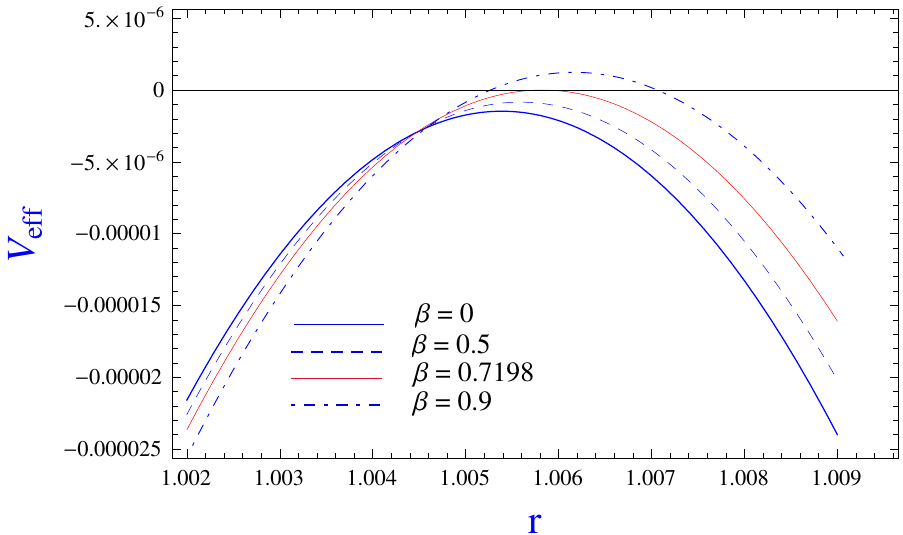}% Here is how to import EPS art

%d)  \includegraphics[width=0.45\textwidth]{fig10.eps} %
\caption{\label{rn_eff} Radial profile of the effective
potential for the radial motion of the charged particles around the Reissner-Nordstr\"{o}m black hole immersed in an external uniform magnetic field for different values of magnetic parameter $\beta$. Note that we select $q=3\times10^{-3}$ and $m=1.8\times10^{-3}$ for test particle's charge and mass, respectively. }
\end{figure}

To analyze the impact of the magnetic parameter $\beta$ on the dynamics of overcharging of the black hole we present the radial profile of the the effective potential in Fig.~\ref{rn_eff}.  From Fig.~\ref{rn_eff}, in the limit $\beta\to 0$ we have the effective potential that is always negative for the charged particle with appropriate parameters, as presented in Table~\ref{table1} and Fig.~\ref{figure2}. However, the shape of the effective potential moves upward as a consequence of the increase in the value of the magnetic parameter $\beta$. As seen in Fig.~\ref{rn_eff}, the height of maximum of
the effective potential tends to zero when the magnetic parameter approaches
its critical value $\beta_{cr}$. Then the maximum value of the effective potential crosses zero and becomes positive, thereby preventing particles from entering the black hole. It turns out that the external magnetic field would act as a cosmic censorship beyond its threshold value; see Fig.~\ref{rn_eff}. 
Table~\ref{table2} also reflects the  behavior of the obtained results in detail.
\begin{figure*}%[pb]
\centering
\includegraphics[width=0.45\textwidth]{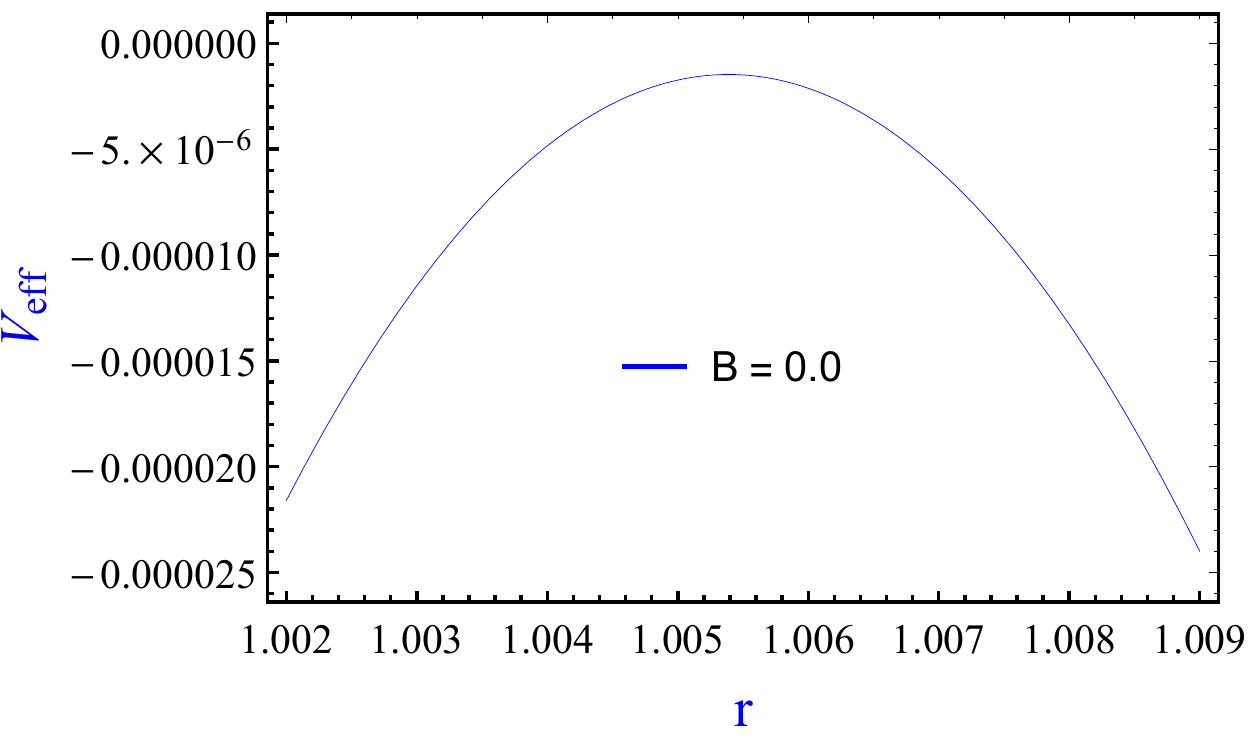}% 
\includegraphics[width=0.45\textwidth]{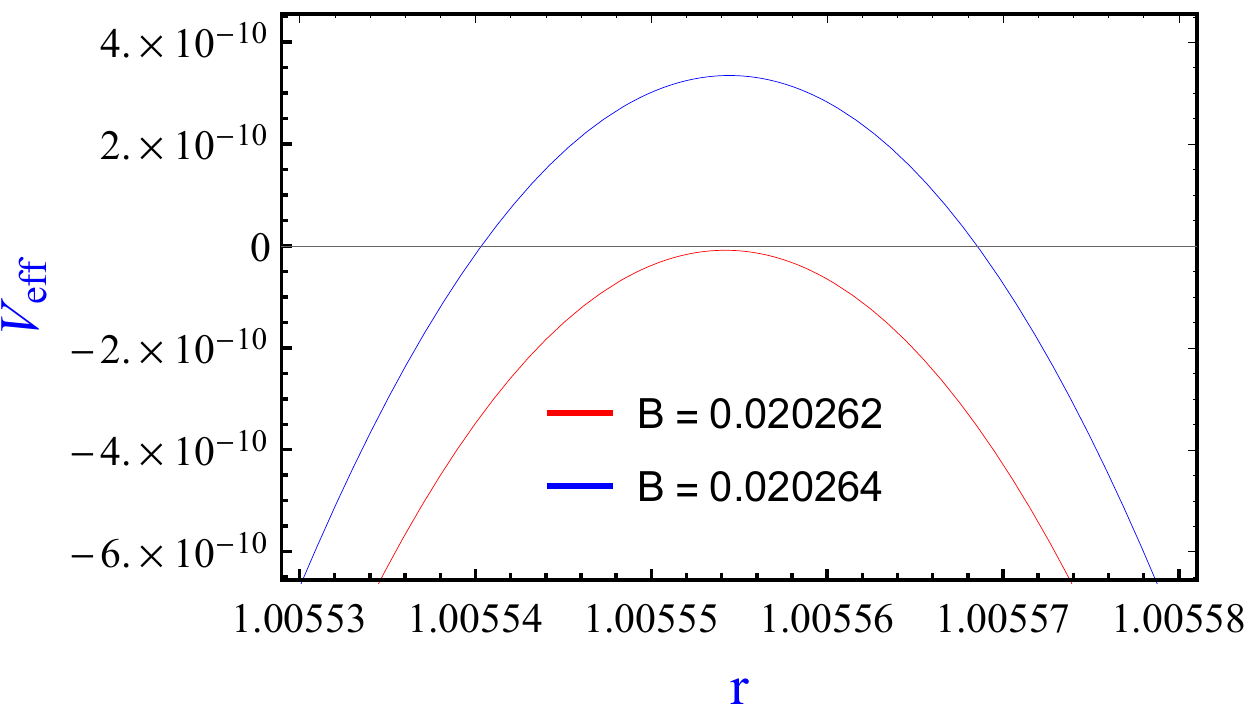} %
 %\vspace*{8pt}
\caption{\label{fig:1} Radial profile of the effective potential for the charged particle around the magnetized Reissner-Nordstr\"{o}m black hole in the case of fixed parameters $\delta E=2.9897\times10^{-3}$, $q=3\times10^{-3}$ and $m=1.8\times10^{-3}$. Left panel: $V_{eff}$ is plotted in the case of vanishing magnetic field, i.e. $B=0$. Right panel: $V_{eff}$ is plotted for different values of magnetic field $B$, red line shows the case when $B=B_{cr}$. Note that in the case of $B>B_{cr}$ the magnetic field can restore the WCCC as the maximum height of $V_{eff}$ crosses horizontal zero line.} \end{figure*}

\begin{table}[ht]
\caption{\label{table2} The values of $r_{max}$ outside the
horizon and effective potential $V_{eff}(r_{max})$ for different values of the magnetic parameter $\beta$ for the fixed $q=3\times10^{-3}$ and $m=1.8\times10^{-3}$.\\} \centering
\begin{tabular}{c c c c }
\hline\hline
% &~  &~ $\tilde{c}=10^{-3}$ &~  &~  \\[0.5ex] \hline
 &~  &~ $\delta E=2.9897\times10^{-3}$ &~   \\[0.5ex] \hline
% \hline
%\multicolumn{3}{c}{Item} \\
%\cmidrule(r){3-5}
$\beta$ &~  &~ $r_{max}$ &~ $V_{eff}$ \\ \\
0 &~  &~ 1.005389149 &~                  $-1.18000\times10^{-6}$  \\
0.01 &~             &~ 1.005389226 &~ $-1.17937\times10^{-6}$ \\
0.05&~   &~ 1.005391085 &~ $-1.17503\times10^{-6}$  \\
0.1 &~  &~ 1.005396903 &~ $-1.16139\times10^{-6}$ \\
0.2 &~  &~ 1.005420304 &~  $-1.10587\times10^{-6}$  \\
0.3 &~  &~ 1.005459767 &~  $-1.00993\times10^{-6}$ \\
0.4 &~  &~ 1.005516007 &~  $-0.86830\times10^{-6}$  \\
0.5 &~  &~ 1.005590070 &~  $-6.73278\times10^{-7}$  \\
0.6 &~  &~ 1.005683387 &~  $-4.14325\times10^{-7}$  \\
0.7 &~  &~ 1.005797845 &~  $-7.78888\times10^{-8}$  \\
0.7198 &~  &~ 1.005823285 &~  0.00000  \\
0.8 &~  &~ 1.005935891 &~  $3.54872\times10^{-7}$  \\
0.9 &~  &~ 1.006100678 &~  $0.90611\times10^{-6}$  \\
1.0 &~  &~ 1.006296273 &~  $1.60574\times10^{-6}$  \\
[1ex] \hline\hline
\end{tabular}
\end{table}

From the above analysis it turns out that the Reissner-Nordstr\"{o}m black hole could be overcharged. This result is however overturned when the external magnetic field around the black hole is taken into account.  That is, the magnetic field can restore the WCCC beyond when it reaches its critical value (for example, see also, \cite{Shaymatov15}). 
However, we have yet to reach the definite conclusion whether the magnetic field still act as a cosmic censor or not. Thus, we need to consider the magnetized black hole solution that includes the magnetic field in the background spacetime~\cite{Gibbons13} as our main purpose is to study the backreaction effect of the magnetic field on the validity of the WCCC. This is what we plan  to address in the next.

\subsection{The effect of magnetic field on dynamics of overcharging of magnetized Reissner-Nordstr\"{o}m black hole}

We have shown that the external magnetic field can potentially serve as the cosmic censor beyond its critical value, preventing particle from entering the Reissner-Nordstr\"{o}m black hole. Here, we analyze the impact of the magnetic field backreaction in the process of overcharging of a magnetized Reissner-Nordstr\"{o}m black hole, as described by the metric~Eq.\ref{Eq:metric}. We further consider a radially falling charged particle, $\delta J=0$, that can violate the WCCC in the absence of magnetic field in the environment of the black hole \cite{Hubeny99}. In doing so, we try to understand whether the backreaction effect of the magnetic field can still serve as the cosmic censor, thus stopping particles with appropriate parameters from entering the black hole.

\textcolor{black}{As was mentioned above we assume that a charged particle that has energy $\delta E\ll M$ and charge $q\ll Q$ remains valid for the test particle approximation.} Then, it adds extra mass and charge to black hole's mass and charge, respectively,  when it gets absorbed by the black hole. A so-called "Overcharged" can be realized if and only if the conditions given by Eqs. (\ref{Eq:11}-\ref{upper}) are hold. 
We then analyze the effective potential so as to understand
more deeply the effect of the magnetic field backreaction on the process of overcharging of a black hole. To ensure that the particle enters the black hole without encountering any turning point the effective potential (\ref{Eff1}) has to be always negative everywhere outside the horizon, i.e. $V_{\rm eff}<0$. 
\begin{table*}
\begin{center}
\tiny
\caption{The values of $r_{max}$ outside the
horizon and effective potential are tabulated in the case of different
values of $\delta E$, $m$ and $B$. Note that the charge of test particle is taken to be $q=3\times10^{-3}$.}\label{table3}
\resizebox{\linewidth}{!}
{
\begin{tabular}{ l |l l l| l l l}
% \begin{tabular}{c c c c | c c | c c | c c}
 \hline \hline
  &\multicolumn{2}{c}{$\delta E=2.9874\times 10^{-3}$}& $m=0.9\times10^{-3}$ & \multicolumn{2}{c}{$\delta E=2.9884\times 10^{-3}$}& $m=1.4\times10^{-3}$ \\
\cline{2-3}\cline{4-7}
 & $B$ & $r_{max}$ & $V_{eff}$ & $B$ & $r_{max}$ & $V_{eff}$  \\
\hline
   & 0        & 1.004630 & $-4.2\times 10^{-6}$ & 0 & 1.004960 & $-2.17\times 10^{-6}$  \\
   & 0.013841 & 1.004681 & 0 & 0.016084 & 1.005045 & 0  \\
   & 0.013843 & 1.004685 & $4.2\times 10^{-10}$ & 0.016085 & 1.005050 & $1.2\times 10^{-10}$ \\
\hline
&\multicolumn{2}{c}{$\delta E=2.9897\times 10^{-3}$}& $m=1.8\times10^{-3}$ & \multicolumn{2}{c}{$\delta E=2.9899\times 10^{-3}$}& $m=1.85\times10^{-3}$ \\
\cline{2-3}\cline{4-7}
 & $B$ & $r_{max}$ & $V_{eff}$ & $B$ & $r_{max}$ & $V_{eff}$  \\
\hline
   & 0 & 1.005390 & $-1.18\times 10^{-6}$ & 0 & 1.005460 & $-1.07\times 10^{-6}$  \\
   & 0.020262 & 1.005555 & 0 & 0.020861 & 1.005640 & 0  \\
   & 0.020263 & 1.005556 & $2.0\times 10^{-10}$ & 0.020862 & 1.005640 & $0.87\times 10^{-10}$ \\
  \hline \hline
\end{tabular}
}
\end{center}
\end{table*}
For that we   assume $M=1$ and $Q=0.99999$ as
stated previously and focus on the charged particle that has
$q=3\times10^{-3}$. Using suitable choice of parameters $\alpha^{\prime}$,
$\beta^{\prime}$, and $\gamma^{\prime}$ given by Eqs.~(\ref{abs})
we find the the appropriate range of energy as
\begin{eqnarray} \label{range_en}2.9866135\times10^{-3}=\delta E_{min}<\delta E< \delta
E_{max}=2.99\times10^{-3}\, . \end{eqnarray}
The particle mass for that can have $m \lesssim 1.96\times10^{-3}$ for given value of  charge $q$.

In Fig.~\ref{fig:1} we show the impact of the magnetic field backraction on the radial profile of the effective potential. From the Fig.~\ref{fig:1}, if the magnetic field is absence there exists parameter space available for test particle that could reach the horizon and lead to overcharging of black hole, i.e. it could violate the WCCC (see, left panel). From the right panel of Fig.~\ref{fig:1}, $V_{\rm eff}$ tends upward as a consequence of the increase in the value of the magnetic field $B$. The charged particle then cannot reach the horizon when $V_{\rm eff}$ reaches the zero at which $B=B_{cr}$. This happens because the parameter space required for overcharging turns out to be closed for particle to reach the horizon.
Thus there is no parameter space available for charged test particles
for violating the WCCC. This is exactly what happens for the magnetized black hole, the threshold value of the magnetic field could lead to restoring the WCCC. In Table~\ref{table3} we show the numerical values of the effective potential for charged test particle in a variety of possible cases. Note that the result shown in Fig.~\ref{fig:1} refers to a particular case of results presented in Table~\ref{table1}.  As shown in Fig.~\ref{fig:1} one can easily notice that $B_{cr}$ corresponds to $V_{eff}=0$ and increases as one increases appropriate parameters of charged test particle; see Table~\ref{table3}. From the above analysis, overcharging test particle would not be able to approach horizon to enter the black hole when the magnetic field reaches its minimum threshold value. Thus, the magnetic field can serve as the cosmic censor, resulting in restoring the CCC in the weak form.

\section{Conclusion}
\label{Sec:Conclusion}

In this paper, we have studied   
the dynamics of overcharging the magnetized Reissner-Nordstr\"{o}m  black hole via the charged test particle 
with appropriate geodesic parameters. It is possible to allow a transition to occur from a Reissner-Nordstr\"{o}m  black hole to Reissner-Nordstr\"{o}m naked singularity. In the astrophysical scenario, it is believed that black holes are surrounded with an external magnetic field that can affect drastically on the charged particle motion due to dominating  Lorentz force. Thus, we have investigated the effect of the test magnetic field on the dynamics of overcharging a black hole. Interestingly we show that it is not possible for a charged particle with the appropriate parameters to violate the WCCC in the case of sufficiently large values of the magnetic field. To understand the effect of magnetic field in detail, we used the solution depicting the magnetized Reissner–
Nordström black hole that involves the effect of the magnetic field on the background geometry. We show that the magnetic field restores the WCCC when it reaches its threshold value. 

The main conclusions of investigation performed are summarized as follows:

\begin{itemize}

\item It is well known that test magnetic field surrounding black hole could give a dominating effect to the motion of
charged test particles due to the Lorentz force. It turns out that a test magnetic field would serve as a cosmic censor, thus being impossible for particles to enter the black hole~\cite{Shaymatov15}. However, the situation is radically different for a black hole immersed in an external test magnetic field in contrast to the exact solution representing the magnetized black hole containing the backreaction of the magnetic field on the background geometry. Thus, the magnetized black hole solution~\cite{Gibbons13} allows to consider the backreaction of the magnetic field both on background geometry and the charged particle motion as well. In order to understand the behavior of the magnetic field,  the process of overcharging a near extremal magnetized Reissner-Nordstr\"{o}m black hole~\cite{Gibbons13} with a charged particle is studied.  

\item We have found that it is also possible for a charged particle to overcharge the magnetized black hole for sufficiently small values (in contrast to its critical value) of the magnetic field. However, it is intriguing that overcharging is controlled by the magnetic field. The magnetic field would therefore restore the cosmic censorship conjecture. This happens because  magnetic field would not allow charged particle to reach the black hole horizon. With this we have shown that the same is true for backreaction of the magnetic field -- it would act as cosmic censor. 

\end{itemize}

\section*{Acknowledgments}
 S.S. and B.A. wish to acknowledge the support from Research Grant  No. F-FA-2021-432 of the Uzbekistan Ministry for Innovative Development and from the Abdus Salam International Centre for Theoretical Physics under the Grant No. OEA-NT-01. 

\section*{Data Availability Statement} This manuscript has no associated data.

\bibliographystyle{apsrev4-1}  %% BibTeX style
\bibliography{gravreferences,Lensing,ref}

\end{document}